\documentclass[twocolumn,showpacs,preprintnumbers,amsmath,amssymb]{revtex4-1}

\usepackage{graphicx}
\usepackage{amsmath}
\usepackage{amsfonts}
\usepackage{amssymb}
\usepackage{pstricks}
 \usepackage{psfrag}
 \usepackage{epsfig}
 \usepackage[ansinew]{inputenc} 
 \usepackage{nicefrac}
 \usepackage{color}

\newcommand{\vecd}[1]{\mathbf{#1}}
\newcommand{\mat}[1]{\mathcal{#1}}
\newcommand{\Wi}{\text{Wi}}
\newcommand{\dev}{\tilde}
\newcommand{\dew}{\bar}

\begin{document}

\title{Dynamics of two trapped Brownian particles:\\shear-induced cross-correlations}

\author{Jochen Bammert, Lukas Holzer, Walter Zimmermann}

\affiliation{
Theoretische Physik I, Universit\"at Bayreuth, D-95440 Bayreuth
}

\date{Received: \today / Revised version: \today}

\begin{abstract}
The dynamics of two Brownian particles trapped by two neighboring harmonic potentials in a linear shear flow is investigated.
The positional correlation functions in this system are calculated analytically and analyzed as a function of the
shear rate and the trap distance.
Shear-induced cross-correlations between particle fluctuations along orthogonal directions in the shear plane are found.
They are linear in the shear rate, asymmetric in time, and occur for one particle as well as between both particles.
Moreover, the shear rate enters as a quadratic correction to the well-known correlations of random displacements along parallel spatial directions.
The correlation functions depend on the orientation of the connection vector between the potential minima with respect to the flow direction.
As a consequence, the inter-particle cross-correlations between orthogonal fluctuations can have zero, one or two local extrema as a function of time.
Possible experiments for detecting these predicted correlations are described.
\end{abstract} 


\maketitle

\section{Introduction}
\label{sec: intro}
The Brownian dynamics of particles in fluids is of high relevance
in many fields of natural and applied sciences. It is
strongly affected by the interplay between the particles via the liquid, the so-called
hydrodynamic interaction.
Especially in the field of microfluidics this nonlinear interaction plays an important role
in subjects such as Taylor dispersion \cite{TaylorGI:1953.1} or fluid mixing \cite{Ottino:2004.1,Steinberg:2001.2}.
In a quiescent fluid there is already a considerable understanding of the dynamics 
of Brownian particles and their interactions \cite{Dhont:96,DoiEd:86}.
Investigations on the positional correlation functions of two trapped particles
give further insight in the coupling between thermal motion and the hydrodynamic interaction among them
\cite{Quake:1999.1,Stark:2004.1}.
However, the understanding of this interplay in typical laminar
flows, like in a linear shear flow or in a Poiseuille flow,
is still far from complete although it is the origin of
a number of interesting phenomena.
For example, polymers exhibit in shear flows already at
small values of the Reynolds number the
so-called molecular individualism \cite{Chu:1999.1,deGennes:1997.1},
elastic turbulence, and spectacular mixing properties
in the dilute regime \cite{Steinberg:2000.1}. 
The complex interplay between  Brownian motion and hydrodynamic interaction also
affects considerably 
the conformational distribution functions 
of tethered polymers in flows and their dynamics
\cite{Chu:1995.1,Brochard:1993.1,Larson:1997.1,Rzehak:99.2,Rzehak:00.1,Kienle:2001.1,Rzehak:2003.2,Cieplak:2006.1,Kienle:2010.1}. 
When polymers are attached to a wall and subjected to a flow, an additional time-periodic behavior influences the dynamics
\cite{Doyle:2000.1,Yeomans:2005.1,Delgado:2006.1,Shaqfeh:2009.1,Graham:2009.1},
which shows similar features as tumbling polymers in shear  flows
\cite{Chu:1999.1,Larson:2000.4,Shaqfeh:2005.2,Steinberg:2006.3}.

Recent theoretical investigations on the
fluid-velocity fluctuations in shear flows show that in contrast to quiescent fluids or uniform flows,
 cross-correlations  between velocity fluctuations 
along and perpendicular to the streamlines occur \cite{Eckhardt:2003.1,Oberlack:2006.1}.
For free Brownian particles in a linear shear flow in $x$ direction, where the shear plane is parallel to the $xy$ plane,
 one also expects a cross-correlation  
between the orthogonal positional fluctuations $\dev{x}$ and $\dev{y}$ of the particles, i.e. $\langle \dev{x}\dev{y} \rangle \neq 0$
\cite{Bedeaux:1995.1,Brady:04,Drossinos:05}. In this case, random jumps of a particle
between neighboring parallel streamlines
lead to a change of the particle's velocity. For example, a positional
fluctuation, $\dev{y}$, perpendicular to the streamlines may cause a fluctuation, $\dev{x}$,
along the streamlines  and contributes in this way to the correlation function
$\langle \dev{x}\dev{y} \rangle$, which reflects the shear-induced coupling between fluctuations 
along orthogonal directions.

The theoretical considerations described in Ref.~\cite{Holzer:2009.1} show that shear-induced
cross-correlations between perpendicular random particle displacements, like 
$\langle \dev{x}\dev{y} \rangle \not =0$, survive   if a particle experiences some constraints
such as a harmonic potential. This is important from various points of view.
First, these cross-correlations
are inherently present in bead-spring models, which are used to describe polymer dynamics in shear flows,
because the individual beads along the chain are bound to their neighbors.
Second, this knowledge
facilitates the experimental detection of these cross-correlations,
because measurements of particle fluctuations in the spatially limited
area of the trapping potential can be performed in a controlled manner compared
to tracing free Brownian particles.
According to this strategy, the cross-correlations $\langle \dev{x}\dev{y} \rangle$ of trapped particles
in a linear shear flow have been measured directly
for the first time and the results are in good
agreement with the theoretical predictions \cite{Ziehl:2009.1}.

The optical tweezer 
technique, employed in Ref.~\cite{Ziehl:2009.1}, triggered a number of further 
direct observations of particle fluctuations.
These include inspiring studies on single polymers
\cite{Chu:1995.1,Brochard:1993.1,Larson:1997.1,Simons:BPJ70-96-1813}, the propagation
of hydrodynamic interactions \cite{Bartlett:2002.1}, wall effects on Brownian motion \cite{Grier:2000.1,Franosch:2009.1},
two-point microrheology
\cite{Weitz:2000.1}, particle sorting
techniques \cite{Grier:2002.1,Dholakia:2003.1,Bammert:2008.1,Bammert:2009.1}, the 
determination of the effective pair potential in colloidal suspensions
\cite{Crocker:JCIS179-96-298}, and many other investigations
in microfluidics, cf. \cite{Florin:2005.1,Schmidt:2005.1}. By femto-Newton measurements 
anti-correlations have been detected 
between two hydrodynamically interacting and 
neighboring particles in a quiescent fluid, each one captured by a
laser-tweezer potential \cite{Quake:1999.1}. This was
recently extended in Ref.~\cite{Ziehl:2009.1}, where 
shear-induced inter-particle anti-correlations between
orthogonal motions of the two particles have been found.

The present work focuses on the question, which kind of 
cross-correlations can be expected
between hydrodynamically interacting particles in linear shear flows.
Such inter-particle correlations along a single polymer and between different polymers in 
flows influence their dynamics.
This paper is an extension of the theoretical work on
 the single particle dynamics described in Ref.~\cite{Holzer:2009.1} 
to a pair of two hydrodynamically interacting point-particles
with an effective hydrodynamic radius and  trapped in a linear shear flow. It
provides the theoretical background for the experimental 
results on the two-particle correlations presented in Ref.~\cite{Ziehl:2009.1}.
The correlation functions between the different particle
displacements are calculated analytically and
we show how a second Brownian particle influences 
the stochastic motion and the positional probability distribution
of its neighbor compared to the
single particle case \cite{Ziehl:2009.1,Holzer:2009.1}.
In addition, we find 
that the anti cross-correlations between two fluctuating 
particles in a quiescent fluid, as described in 
Ref.~\cite{Quake:1999.1}, experience shear-induced corrections.
We also describe the occurrence of shear-induced anti cross-correlations
between the fluctuations of the two particles
along two orthogonal directions.
The results depend significantly on the orientation of the
connection vector between the two traps with respect to the flow 
direction.
We focus on the leading order contributions to
the correlation functions and neglect the effects
of finite size and rotations of the particles.

The structure of the paper is as follows:
In sect.~\ref{sec: model}, the model equations are introduced and
their formal analytical solution is presented, which results
in the calculation of the correlation functions for
the random particle displacements.
In sect.~\ref{sec: corr},
the results for three representative setups of
the two-particle system are discussed in detail, where
the connection vector between the two potential minima is
either parallel, perpendicular, or oblique to the shear-flow direction.
In addition, the results
are compared with direct simulations of the Langevin equation
for representative examples.
For the parallel case, a few experimental and theoretical
results have already been described  
in Ref.~\cite{Ziehl:2009.1},
 where a good agreement between experiment
and theory was found. The article closes 
with a discussion and further possible applications in sect.~\ref{sec: conclusion}.

\section{Equations of motion and their solution}\label{sec: model}

The basis of our investigation is a Langevin model that
describes the over-damped dynamics of two particles, each held
by a harmonic potential in a linear shear flow. 
In this section the equation of motion is introduced and solved in
order to calculate the correlation functions analytically.
They consist of different eigenmodes, which are discussed briefly.

\subsection[Model]{Model equations}\label{sec: eq}

\begin{figure}
  \begin{center}
 \includegraphics[width=0.7\columnwidth]{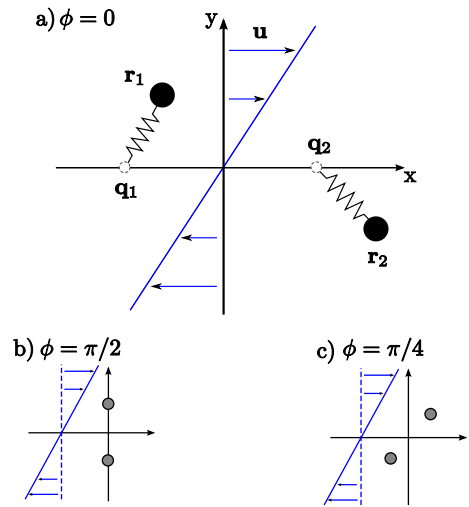}
  \end{center}
\caption{Two Brownian particles are kept by linear spring-forces
$\vecd{f}_{1,2}^V$ close to the minima of
two corresponding harmonic potentials at $\vecd{q}_{1,2}$.
Both particles are simultaneously exposed to a linear shear flow
$\vecd{u}(\vecd{r})$.
The angle $\phi$ between the 
flow direction $\vecd{u}(\vecd{r})$ and the vector
$\vecd{q}_{12}=\vecd{q}_1-\vecd{q}_2$ connecting
the potential minima at the distance $ d= |\vecd{q}_{12}|$ is either zero, as in part
a), or $\phi=\pi/2$ as in b), or $\phi=\pi/4$ as in c).}
\label{fig:01}
\end{figure}

We consider two Brownian point-particles  with 
effective hydrodynamic radius $a$, immersed in a Newtonian fluid 
of viscosity $\eta$ at the 
positions 
$\vecd{r}_i=(x_i,y_i,z_i)$ with $i=1,2$. 
Each particle is held by a linear
restoring force,
\begin{align}
\label{fv_eq}
 \vecd{f}^V_{i}=-\nabla V_{i}=-k\left(\vecd{r}_{i}-\vecd{q}_i \right)\,,
\end{align}
close to the  minimum $\vecd{q}_i$ 
of a corresponding harmonic potential,
\begin{align}
\label{pot_eq}
 V_i=\frac{k}{2}\left(\vecd{r}_i-\vecd{q}_i \right)^2\,,
\end{align}
with the spring constant  $k$. The distance between the potential minima is labeled with $d$.
Uncharged polystyrene latex beads of micrometer size that are
trapped by laser tweezers experience such a potential as described by 
eq.~(\ref{pot_eq}) \cite{Quake:1999.1,Chu:1991.1,Grier:2008.1}.

Both trapped particles
are exposed to a linear shear flow in the $x$ direction with the 
shear plane parallel to the $xy$ plane and the shear rate $\dot{\gamma}$:
\begin{align}
\label{u_eq}
\vecd{u}(\vecd{r})=\dot{\gamma}y\vecd{e}_x\,.
\end{align}
This flow causes a drag force on the hydrodynamically interacting Brownian particles,
which is in competition with the restoring force (\ref{fv_eq}).
In a recent experiment
the interplay between these two forces
has been studied \cite{Ziehl:2009.1}, where
the size of the particles was about $5\mu m$ and 
the shear rate about $50 s^{-1}$.
At the length scale of the excursions
 of the
particles from their potential minima and
the distance 
 between the two beads, the Reynolds number
is small and therefore, we can describe the fluid
motion around the beads in the Stokes limit. Consequently, the
over-damped particle dynamics is described
by the Langevin equations,
\begin{align}
 \label{dgl_eq}
 \dot{\vecd{r}}_i =\vecd{u}(\vecd{r}_i) +
{\mat{H}}_{ij} \vecd{f}^V_j  + \vecd{f}^S_i  \,,
\end{align}
with the four $3\times3$ mobility matrices
\begin{subequations}
\begin{align}
\label{stokes_eq}
  {\mat{H}}_{11}& = {\mat{H}}_{22}=\frac{1}{\zeta}\mat{I}\,, \\
\label{oseen_eq}
  {\mat{H}}_{12}& = {\mat{H}}_{21}= \frac{1}{\zeta} \frac{3a}{4 r_{12}}
\left[{\mat{I}} + \frac{\vecd{r}_{12} \otimes
\vecd{r}_{12}}{r_{12}^2}\right] \,,
\end{align} 
\label{Hij_eq}
\end{subequations}
including the Stokes friction coefficient $\zeta=6 \pi \eta a$ for
a single point-particle.
 \cite{Stokes:1850.1}.
The latter two matrices describe the hydrodynamic interaction between
two point-particles in terms of the Oseen tensor
\cite{DoiEd:86} and $\mat{I}$ represents the unity matrix.
The dyadic product (tensor product)  $\otimes$ has been used as well as the distance
vector $\vecd{r}_{12}:=\vecd{r}_1-\vecd{r}_2$ 
between the beads, with $r_{12}=|\vecd{r}_{12}|$.

The stochastic forces in the Langevin model have their origin
in the velocity fluctuations of the surrounding liquid. In
a quiescent fluid these random forces are uncorrelated along orthogonal
directions in the bulk \cite{LanLifVI}.
This assumption is kept in our Langevin model, because shear-induced
cross-correlations between the stochastic forces
along orthogonal directions are expected to be small \cite{Oberlack:2006.1,Holzer:2009.2,Rzehak:2003.1}.
So, for the contribution $\vecd{f}^S_i(t)$ in eq.~(\ref{dgl_eq}) we assume a zero mean and a
vanishing correlation time \cite{DoiEd:86}:
\begin{subequations}
\begin{align}
 \label{stoch_eq1}
 \langle  \vecd{f}^S_i(t) \rangle &  = 0\,, \\
 \label{stoch_eq2}
 \langle  \vecd{f}^S_i (t) \otimes \vecd{f}^S_j(t')\rangle &  = 2k_BT  {\mat{H}}_{ij} \delta(t-t')  \,.
\end{align}
\label{stoch_eq}
\end{subequations}
The strength of the stochastic forces is proportional
to the thermal energy $k_BT$.

The orientation of the connection vector between the potential minima, 
$\vecd{q}_{12}:=\vecd{q}_1-\vecd{q}_2$, 
with respect to the flow direction is described
by the angle $\phi$.
It has a strong influence on the
correlation functions of the positional fluctuations
of the particles. For this reason,
we investigate three characteristic setups, where
$\vecd{q}_{12}$ is either parallel to the external flow $\vecd{u}$ ($\phi=0$), or
perpendicular ($\phi=\pi/2$), or oblique ($\phi=\pi/4$) as sketched in fig.~\ref{fig:01}.

\subsection{Solutions and relaxation times}\label{sec: calc}

There are two characteristic time scales in the system. One is 
determined by the inverse shear rate $\dot{\gamma}^{-1}$ and the other one is 
given by the relaxation time  $\tau:=\zeta/k$ of the particles
in the two identical potentials. 
Their ratio gives the dimensionless Weissenberg number,
\begin{align}
 \label{wi_eq}
\Wi := \dot{\gamma}\tau\,,
\end{align}
which will be useful for the further discussion.

The first step in the solution of eq. (\ref{dgl_eq})
is to rewrite the equation of motion in a more compact
form by introducing the positional vector
$ \vecd{R}=(\vecd{r}_1,\vecd{r}_2)$
with six components
and the $6\times 6$ mobility matrix $\mat{H}$,
\begin{align}
 \label{matH6_eq}
 \mat{H}=\begin{pmatrix} \mat{H}_{11} &    \mat{H}_{12} \\ \mat{H}_{12} &    \mat{H}_{22}
         \end{pmatrix}\,,
\end{align}
composed of the 
sub-matrices ${\mat{H}}_{ij}$ from eqs. (\ref{Hij_eq}).
The shear flow in eq.~(\ref{u_eq}) can be written in an analogous manner
with the $6\times 6$ shear rate tensor $\mat{U}$,
\begin{align}
\label{U_eq}
  \vecd{U}(\vecd{R})=\mat{U} \vecd{R}\,,
\end{align}
with ${\mat{U}}_{12}={\mat{U}}_{45}=\dot{\gamma}$ and all other ${\mat{U}}_{kl}=0$.
The equation of motion (\ref{dgl_eq})
then takes the form,
\begin{align}
 \label{dgl2_eq}
 \dot{\vecd{R}}&={\mat{U}} \vecd{R}+k {\mat{H}} \left(\vecd{Q}- \vecd{R}\right)+\vecd{F} \,,
\end{align}
where the vector $\vecd{Q}=(\vecd{q}_1,\vecd{q}_2)$ describes the positions of the
two potential minima being separated by the distance $d$. 
The stochastic contribution $\vecd{F}$ in  eq.~(\ref{dgl2_eq}) is obtained from eqs.~(\ref{stoch_eq}):
\begin{subequations}
\begin{align}
 \label{stoch2_eq1}
   \langle \vecd{F}(t) \rangle &   = 0\,, \\
 \label{stoch2_eq2}
 \langle \vecd{F}(t) \otimes \vecd{F}(t') \rangle &   = 2k_BT {\mat{H}} \delta(t-t')  \,.
\end{align}
\label{stoch2_eq}
\end{subequations}
We assume two well separated point-particles with small values of $a/d$ and
small fluctuations around their mean positions, i.e. $k_BT/(ka^2) \ll 1$.
The experimental results described in Ref.~\cite{Ziehl:2009.1}, which  were obtained for
$a/d \approx 1/4$ and a magnitude of the fluctuations below $a/10$, are well described 
within this approximation.

Since we investigate the particle fluctuations, the mean position $\vecd{R}_\phi := \langle \vecd{R}(t) \rangle$
has to be determined first.
In the case $\phi=0$ the mean positions are identical
with the locations of the potential minima: $\vecd{R}_0=\vecd{Q}$. 
For $\phi=\pi/2$ and $\phi=\pi/4$, $\vecd{R}_\phi$ is obtained numerically
by determining the stationary solution of
 eq.~(\ref{dgl2_eq}) in the absence of noise.
Disregarding the hydrodynamic interaction between the two particles 
one finds the analytical expressions,
\begin{subequations}
\begin{align}
\vecd{R}^a_{\pi/2}& =\frac{d}{2}\left(\Wi,1,0,-\Wi,-1,0\right)\,, \\
\vecd{R}^a_{\pi/4}& =\frac{d}{4}\left(\sqrt{2}+\Wi,\sqrt{2},0,-\sqrt{2}-\Wi,-\sqrt{2},0\right)\,, 
\end{align}
\label{meansenk_eq}
\end{subequations}
which may serve as an approximation of the stationary solution 
and as the starting point of the numerical iteration.

The equation of motion for the particle fluctuations $\dew{\vecd{R}}=(\dew{x}_1,\dew{y}_1,\dew{z}_1,\dew{x}_2,\dew{y}_2,\dew{z}_2)$
are obtained by the ansatz $\vecd{R}=\vecd{R}_\phi+\dew{\vecd{R}}$ and the linearization of
eq.~(\ref{dgl2_eq}) with respect to  $\dew{\vecd{R}}$:
\begin{align}
 \label{dgl3_eq}
\dot{\dew{\vecd{R}}} = \mat{U}\dew{\vecd{R}} -k\mat{H}\dew{\vecd{R}} + k\left[ \nabla \mat{H} \dew{\vecd{R}}\right](\vecd{Q}-\vecd{R}_\phi) + \vecd{F}.
\end{align}
Here the mobility matrix  $\mat{H}$
 and its derivative are evaluated at the exact mean positions $\vecd{R}_\phi$.
For $\phi=0$ one has $\vecd{R}_\phi=\vecd{Q}$ and the third contribution on the right hand side 
vanishes.
Introducing the matrix $\mat{K}:=\left[ \nabla \otimes (\mat{H}(\vecd{Q}-\vecd{R}_\phi))\right]^T$,  eq.~(\ref{dgl3_eq}) can be rewritten to
\begin{align}
 \label{dgl14_eq}
\dot{\dew{\vecd{R}}} = - \left( k\mat{H} - \mat{U} - k\mat{K} \right)\dew{\vecd{R}} + \vecd{F} = -\mat{M} \dew{\vecd{R}} + \vecd{F} \,,
\end{align}
and this linear equation has the formal solution:

\begin{align}
 \label{rt_eq}
 \dew{\vecd{R}}(t)=e^{-t{\mat{M}}}\dew{\vecd{R}}(0)+\int_0^t dt' e^{(t'-t){\mat{M}}} \vecd{F} \,.
\end{align}
By introducing the scaled deviation
\begin{align}
\label{B_eq}
\dev{\vecd{R}}= \frac{\dew{\vecd{R}}}{\sqrt{B}}\,, \quad \mbox{with} \quad  B=\frac{2k_BT}{k}\,,
\end{align}
and taking into account the
statistical properties of the stochastic forces as
 given by eq.~(\ref{stoch2_eq}),
one can determine by a straight-forward calculation,
assisted by computer algebra,
the correlation matrix ${\mat{C}}(t)$ defined by,
\begin{align}
 \label{corrmat_eqA}
 {\mat{C}}(t) := \langle \dev{\vecd{R}}(0) \otimes \dev{\vecd{R}}(t) \rangle\, \quad \mbox{for}
\quad t\geq0\,.
\end{align}
The brackets $\langle\cdot\rangle$ denote the ensemble average over a large number of particle trajectories.
The  elements ${\mat{C}}_{kl}(t)$ of the Matrix $\mat{C}(t)$ can be represented
as a sum of six exponentially decaying 
contributions,
\begin{align}
\label{ckl_eq}
{\mat{C}}_{kl}(t)=\sum_\alpha g_{\alpha,kl} e^{-\lambda_\alpha t}\quad (\alpha,k,l=1...6)\,,
\end{align}
where the coefficients $g_{\alpha,kl}$ depend on the Weissenberg number
$\Wi=\dot{\gamma}\tau$ and on the distance between the potential minima $d$.

\begin{figure}
  \begin{center}
\includegraphics[width=0.95\columnwidth]{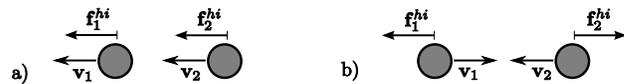}
  \end{center}
\caption{The motion of particle 1 causes via the hydrodynamic interaction a force $\vecd{f}^{hi}_2$ on particle 2 and vice versa.
In part a), where the two particles move in the same direction, the resulting hydrodynamic forces accelerate the motion.
This situation corresponds to the parallel relaxation.
Part b) shows the anti-parallel case, where the forces $\vecd{f}^{hi}_i$ decelerate the motion.
}
\label{fig:02}
\end{figure}

The origin of the relaxation times $1/Re(\lambda_\alpha)$, 
given by the eigenvalues $\lambda_\alpha$ of the
matrix $\mat{M}:=k\mat{H}-\mat{U}-k\mat{K}$, can be explained as follows:
After a stochastic kick that pushes the particles away from their mean positions,
the potential forces start to pull them back.
During this relaxation the particle motion can be decomposed into 
parallel or anti-parallel translations as illustrated in fig.~\ref{fig:02}.
It is the hydrodynamic interaction between the particles, which accelerates or damps
 this process, since the resulting hydrodynamic forces depend on the relative 
particle motions. The beads relax faster, if they move in the same direction.
The whole relaxation process is described by six relaxation rates, $\lambda_\alpha$, two for each
spatial direction and in the two
cases $\phi=\pi/2$ and $\phi=\pi/4$ some of them may even be complex.
 For the configuration $\phi=0$, cf.
fig.~\ref{fig:01} a), two relaxation modes coincide, so there are only four instead of
 six different relaxation times.

\section{Correlation functions}\label{sec: corr}

In the two-particle system, 
the fluctuation statistics of one particle is influenced by its neighbor.
We call the corresponding correlation functions `one-particle correlations' and 
for the inter-particle cross-correlations between the random displacements of two different particles
we use the notation `inter-particle correlations'.

In a previous study on hydrodynamic interactions 
between two trapped Brownian particles in a quiescent fluid,
anti cross-correlations between their random motions along parallel
spatial directions were found \cite{Quake:1999.1}.
We show, that the shear flow alters these correlations and
additionally induces inter-particle cross-correlations
along orthogonal directions in the shear plane.

In this section the exact expressions and the approximations of the 
 correlation functions $\mat{C}_{kl}(t)$, as defined by eq.~(\ref{corrmat_eqA}),
are discussed in detail and they are also compared with numerically obtained solutions 
of the Langevin equation (\ref{dgl2_eq}).
The behavior of $\mat{C}_{kl}(t)$ depends on the trap distance $d$,
and in the limit $d \to \infty$, our formulas become identical to the recently presented
results for a single trapped
particle in a linear shear flow \cite{Holzer:2009.1}. 
Since our results depend on the angle $\phi$ between the connection vector $\vecd{q}_{12}$
and the flow direction
the three characteristic configurations, as sketched in fig.~\ref{fig:01}, are analyzed.

\subsection{Parallel case: $\phi=0$}\label{p0_sec}

At first, we consider the two-particle configuration
with the connection vector $\vecd{q}_{12}$ 
parallel to the flow lines $\vecd{u}$ as sketched in fig.~\ref{fig:01}~a), e.g. $\vecd{Q}=d/2(1,0,0,-1,0,0)$ . 
The discussion of the one-particle 
correlations in sect. \ref{singleA} is
complemented by the analysis of the inter-particle
correlations in sect. \ref{interpartA}.

Similar to the case of two trapped particles
in a quiescent fluid in Ref. \cite{Quake:1999.1} 
there are four different relaxation rates
describing the four relaxation times in the system, cf. sect.~\ref{sec: calc}:
\begin{subequations}
\begin{align}
 \label{ew1_eq}
 \lambda_1 &  =\frac{1+2\mu}{\tau},  & \lambda_3 =\frac{1-2\mu}{\tau},\\
  \label{ew2_eq}
 \lambda_2 &  =\frac{1+\mu}{\tau},~  & \lambda_4 =\frac{1-\mu}{\tau}\,\,.
 \end{align}
\label{ew3_eq}
\end{subequations}
The parameter $0< \mu:=3a/(4d) <3/8$ is a measure for the distance
between the traps.
$\lambda_1$ and $\lambda_3$ correspond to the particle motions parallel and 
anti-parallel to the connection vector $\vecd{q}_{12}$ (longitudinal displacements),
while $\lambda_2$ and  $\lambda_4 $ belong 
to the particle relaxations
perpendicular to $\vecd{q}_{12}$ (transversal displacements).

\subsubsection{One-particle correlations}
\label{singleA} 

\begin{figure}
  \begin{center}
\includegraphics[width=0.95\columnwidth]{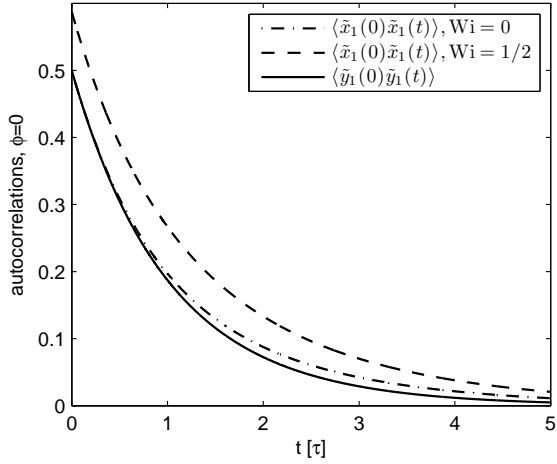}
  \end{center}
\caption{The autocorrelations, $\langle \dev{x}_1(0)
\dev{x}_1(t)\rangle=\langle \dev{x}_2(0) \dev{x}_2(t)\rangle$, along the flow direction
 are shown for $\Wi=0$ (dash-dotted line) and for $\Wi=1/2$ (dashed line).
 The correlation functions perpendicular to the flow direction,
$\langle \dev{y}_i(0) \dev{y}_i(t)\rangle=\langle \dev{z}_i(0) \dev{z}_i(t)\rangle$ with $i=1,2$  
(solid line), do not depend on the Weissenberg number.
All curves are obtained for a distance $d=4a$ and $t$ is given in units of $\tau$.
}
\label{fig:03}
\end{figure}

The autocorrelations are identical for both particles but they are
different for the longitudinal and transversal displacements: $\langle \dev{x}_1(0)\dev{x}_1(t)\rangle = \langle \dev{x}_2(0)\dev{x}_2(t)\rangle$ and 
$\langle \dev{y}_1(0)\dev{y}_1(t)\rangle = \langle \dev{y}_2(0)\dev{y}_2(t)\rangle = \langle \dev{z}_1(0)\dev{z}_1(t)\rangle = \langle \dev{z}_2(0)\dev{z}_2(t)\rangle$.
The expressions
\begin{subequations}
 \label{xx1_0_eq}
\begin{align}
\label{x1x1_0_eq}
  \langle \dev{x}_1(0) & \dev{x}_1(t) \rangle = \frac{1}{4}\left( e^{-\lambda_1 t} + e^{-\lambda_3
t} \right) \nonumber\\
 &  + \frac{\Wi^2}{4\mu}\left( \frac{-(1+\mu)e^{-\lambda_1 t}}{6\mu^2+7\mu+2} 
	+\frac{e^{-\lambda_2 t} }{2+3\mu}\right) \nonumber\\
 &  + \frac{\Wi^2}{4\mu}\left(\frac{(1-\mu)e^{-\lambda_3 t}}{6\mu^2-7\mu+2}
-\frac{e^{-\lambda_4 t}}{2-3\mu}  \right)\,,\\
\label{y1y1_0_eq}
  \langle \dev{y}_1(0) & \dev{y}_1(t) \rangle  = \langle \dev{z}_1(0) \dev{z}_1(t)
\rangle=\frac{1}{4}\left(e^{-\lambda_2 t} + e^{-\lambda_4 t}\right)\,,
\end{align}
\end{subequations}
are exponentially decaying in time and both functions are plotted in fig.~\ref{fig:03} for
different values of the Weissenberg number $\Wi$.
Due to the scaling (\ref{B_eq}) the value of $\langle \dev{y}_1(0)\dev{y}_1(0)\rangle$ is $1/2$.
The correlation functions (\ref{xx1_0_eq}) 
depend via $\mu$ on the trap
distance $d$ , which leads to interesting corrections to the autocorrelations compared to 
the case of one isolated particle in Ref.~\cite{Holzer:2009.1}.

The distinct relaxation rates given by eqs.~(\ref{ew3_eq})
lead to different autocorrelations of particle displacements 
along  and perpendicular to $\vecd{q}_{12}$, independent of the
parameter $\Wi$.
This is also indicated in fig.~\ref{fig:03}
by the difference between the correlations
$\langle \tilde x_1(0) \tilde x_1(t) \rangle$ (dash-dotted line) 
and $\langle \tilde y_1(0) \tilde y_1(t) \rangle$ (solid
line). The latter one is independent of the Weissenberg number, which
is similar to the case of an isolated trapped particle, 
where only the autocorrelation in flow direction
(\ref{x1x1_0_eq}) depends on $\Wi^2$ \cite{Holzer:2009.1}.

\begin{figure}
  \begin{center}
\includegraphics[width=0.95\columnwidth]{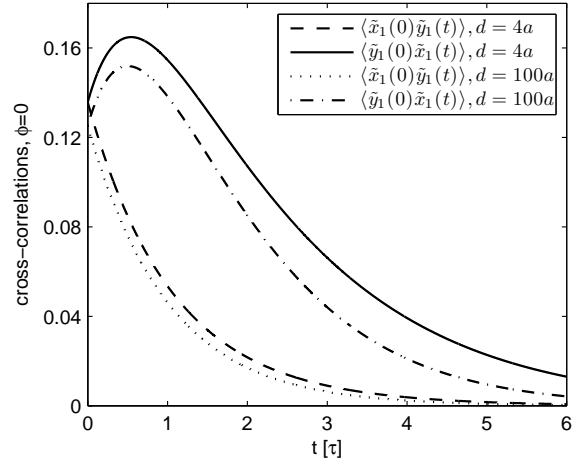}
  \end{center}
\caption{Shear-induced cross-correlations of a single particle in the parallel case for $\Wi=1/2$ and
$d=4a$ respectively $d=100a$. $\langle \dev{x}_1(t)\dev{y}_1(0)\rangle$ has a maximum around $t' \approx \tau$, which 
does not depend on $\Wi$ and only very weakly on $d$.}
\label{fig:04}
\end{figure}

Any translation of a particle in 
$y$ direction is coupled via the flow profile (\ref{u_eq}) to a change
of the particle's velocity in $x$ direction, which leads to a
change of the particle's positional fluctuation along the $x$ direction.
Consequently, the particle fluctuations in the $x$ and $y$ directions become correlated 
due to the shear flow. 
The resulting cross-correlations between the
displacements along orthogonal directions in the shear plane 
are linear functions of the parameter $\Wi$, similar to the single particle case in
Refs.~\cite{Holzer:2009.1,Ziehl:2009.1}.
However, compared to these results, we obtain for the two-particle system 
an additional dependence on the trap distance $d$.
The time-dependence of the cross-correlations 
is given by the following expressions:
\begin{subequations}
 \label{xy1_0_eq}
\begin{align}
\label{x1y1_0_eq}
 \langle \dev{x}_1(0) & \dev{y}_1(t) \rangle = \frac{\Wi}{4}\left( \frac{e^{-\lambda_2 t}}{2+3\mu}
+ \frac{e^{-\lambda_4 t}}{2-3\mu} \right)\,,\\
\label{y1x1_0_eq}
 \langle \dev{y}_1(0) & \dev{x}_1(t) \rangle = \frac{\Wi}{4\mu}\left( e^{-\lambda_2 t} -
e^{-\lambda_4 t} \right) \nonumber \\
 &  + \frac{\Wi}{2\mu}\left(  \frac{(1-\mu)e^{-\lambda_3 t}}{2-3\mu} -
\frac{(1+\mu)e^{-\lambda_1 t}}{2+3\mu}  \right)\,.
\end{align}
\end{subequations}
Both functions are plotted in fig.~\ref{fig:04} for two different
distances. The two sets of curves indicate that the magnitude of the 
cross-correlation increases weakly with
decreasing values of $d$.
The cross-correlation functions involving the $z$ coordinate vanish as in Ref.~\cite{Holzer:2009.1}.

The time-asymmetry of the shear-induced cross\--corre\-lations, namely $\langle \dev{y}_1(0)\dev{x}_1(t)\rangle \neq \langle \dev{x}_1(0)\dev{y}_1(t)\rangle$  with $t>0$, can be explained in the following way:
A random particle displacement at $t=0$ in $y$ direction leads immediately after the kick to a
linear growth of the particle's $x$ coordinate due to the larger flow velocity $\vecd{u}(\vecd{r})$ at
a larger value of the $y$ coordinate.
Consequently, for small values of $t$, the product $\dev{y}_1(0)\dev{x}_1(t)$ grows in time until the
particle is pulled back by the linear spring force, which happens on the time scale  $ \tau=\zeta/k$.
The result is a maximum in the correlation function $\langle \dev{y}_1(0)\dev{x}_1(t)\rangle$ at a time of
the order of $\tau$.
Considering the effect of a random kick in $x$ direction at $t=0$, the particle does not
jump between streamlines of different velocity and therefore 
the cross-correlation $\langle \dev{x}_1(0) \dev{y}_1(t) \rangle$ does not show this maximum and
decays exponentially.

By replacing 
the time $t$ by $-t$ in eq.~(\ref{x1y1_0_eq}) the two functions (\ref{x1y1_0_eq}) and (\ref{y1x1_0_eq}) can be combined to one
correlation function $\langle \dev{x}_1(0) \dev{y}_1(t) \rangle$, where $t$ can now  take  positive
and negative values. This function is asymmetric with respect to time-reflections $t \to -t$.
A similar behavior was previously found for the fluctuations of the fluid-velocity in
a shear flow \cite{Eckhardt:2003.1}.

The static correlation
functions given by eqs.~(\ref{x1x1_0_eq}) -- (\ref{y1x1_0_eq}) for $t=0$ 
determine also the positional distribution function $P(\vecd{r})$ of a Brownian particle in a potential as described in more detail in
Ref.~\cite{Holzer:2009.1}. It is an interesting question, how the
single particle distribution in a shear flow is changed by the
presence of a second one.

\begin{figure}
  \begin{center}
\includegraphics[width=0.95\columnwidth]{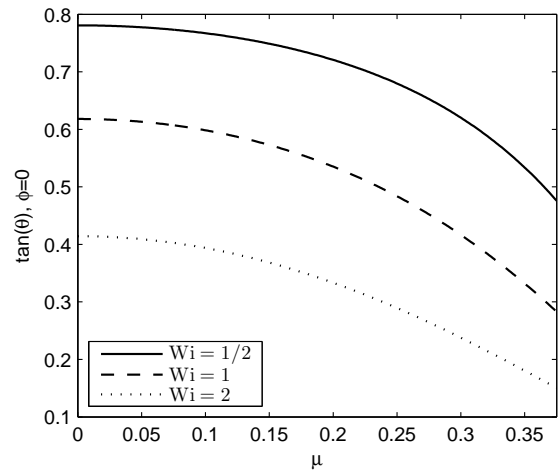}
  \end{center}
\caption{ The inclination angle $\theta$ of the single particle distribution is
shown as function of $\mu=3a/(4d)$ for three different values of the Weissenberg number
$\Wi$.}
\label{fig:05}
\end{figure}

In a linear shear flow $P(\vecd{r})$ has an elliptical shape in the
shear plane and the angle  $\theta$ 
between the major axis of the ellipse and the flow direction is determined by 
the equation
\begin{align}
\label{tanalpha_eq}
\tan \theta & =  \frac{1}{2} \left[ \frac{\langle {\dev{x}_1} {\dev{y}_1}\rangle  }{|\langle
{\dev{x}_1} {\dev{y}_1}\rangle |} \sqrt{  4 + G^2}  - G        \right]\,,\\
\label{tanG}
\mbox{with}&   \qquad G= \frac{\langle {\dev{x}_1}^2 \rangle - \langle {\dev{y}_1}^2 \rangle}{ 
\langle {\dev{x}_1} {\dev{y}_1}\rangle  }\,.
\end{align}
Using the $d$-dependent static correlations from eqs.~(\ref{xx1_0_eq}) and (\ref{xy1_0_eq}) we obtain
\begin{align}
 G=\Wi~\frac{1+3\mu^2}{1-4\mu^2}\,.
\end{align}
Consequently the inclination angle $\theta$ 
is a function of the parameter $d$ too.
In fig.~\ref{fig:05} $\tan(\theta)$ is shown as a function of $\mu=3a/(4d)$ for three
different values of the Weissenberg number $\Wi$ and in all three cases 
$\tan(\theta)$ decreases considerably when the two particles approach. 
Since $\theta$ changes in the same manner 
by decreasing $d$ or increasing $\Wi$, the shear-flow effects 
can be considered to be 
amplified by the presence of the second particle.

\subsubsection{Inter-particle correlations}
\label{interpartA}

\begin{figure}
  \begin{center}
\includegraphics[width=0.95\columnwidth]{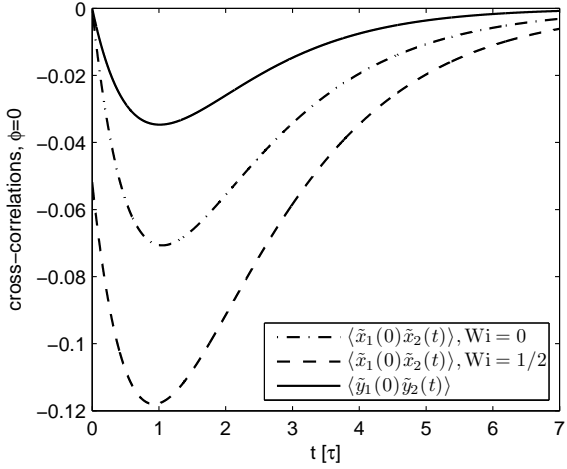}
  \end{center}
\caption{The cross-correlation functions between the two beads along parallel directions are shown for 
$\phi=0$ and $d=4a$. The minimum of $\langle \dev{x}_1(0)
\dev{x}_2(t)\rangle=\langle \dev{x}_2(0) \dev{x}_1(t)\rangle$ increases with the 
Weissenberg number $\Wi$ and is always deeper than the one of the function
$\langle \dev{y}_1(0) \dev{y}_2(t)\rangle=\langle \dev{z}_1(0) \dev{z}_2(t)\rangle$,
which is independent of $\Wi$.}
\label{fig:06}
\end{figure}

The motion of the two trapped Brownian particles is coupled via the
hydrodynamic interaction. In a quiescent fluid this coupling leads to a cross-correlation
between the thermal fluctuations of the two particles 
along the same direction \cite{Quake:1999.1}. Since this cross-correlation
is negative as a function of time, cf. fig.~\ref{fig:06}, 
the notion `anti cross-correlation'
is used. For the anti cross-correlation of the longitudinal displacements we obtain 
a shear-induced correction,  similar as for the one-particle
autocorrelation in eq.~(\ref{x1x1_0_eq}), which is
proportional to $\Wi^2$:
\begin{align}
\label{x1x2_0_eq}
 &  \langle \dev{x}_1(0) \dev{x}_2(t) \rangle= \frac{1}{4}\left( e^{-\lambda_1 t} - e^{-\lambda_3
t} \right) \nonumber\\
  &  + \frac{\Wi^2}{4\mu}\left( \frac{-(1+\mu)e^{-\lambda_1 t}}{6\mu^2+7\mu+2} +
\frac{e^{-\lambda_2 t}}{2+3\mu} \right)\,, \nonumber\\
 &  + \frac{\Wi^2}{4\mu}\left(\frac{-(1-\mu)e^{-\lambda_3 t}}{6\mu^2-7\mu+2}
+\frac{e^{-\lambda_4 t}}{2-3\mu}  \right)\,.
\end{align}
The cross-correlations between random particle-dis\-place\-ments 
perpendicular to $\vecd{q}_{12}$ are independent of
the Weissenberg number $\Wi$:
\begin{align}
\label{y1y2_0_eq}
 &  \langle \dev{y}_1(0) \dev{y}_2(t) \rangle= \langle \dev{z}_1(0) \dev{z}_2(t)
\rangle=\frac{1}{4}\left(e^{-\lambda_2 t} - e^{-\lambda_4 t}\right)\,.
\end{align}
Both correlation functions are plotted in fig.~\ref{fig:06}.

As indicated in fig.~\ref{fig:02}, the relaxation processes of the displacements of
the two particles along the same spatial direction  can be decomposed into a parallel and an anti-parallel translation.
In the present case  the eigenvalue $\lambda_1$ ($\lambda_2$) in eqs.~(\ref{ew3_eq}) corresponds to the parallel relaxation modes along (perpendicular to) the connection vector $\vecd{q}_{12}$.
For the parallel motions, the signs of the particle displacements are always equal for both particles ($+,+$ or $-,-$), whereas for the anti-parallel ones the corresponding displacements have opposite signs ($+,-$ or $-,+$). As a consequence the product of the two displacements is always positive for the parallel case and negative for the anti-parallel case, as indicated by the prefactors of the corresponding contributions in the correlation functions in eq.~(\ref{x1x2_0_eq}) and eq.~(\ref{y1y2_0_eq}).
As the magnitudes of the anti-parallel translations are larger than the parallel ones, the superposition of the two different modes is negative and has a pronounced minimum at a time $t' \approx \tau$, as shown in fig.~\ref{fig:06}.
The value of $t'$ in case of eq.~(\ref{x1x2_0_eq})
 depends only weakly on the distance $d$ between the two traps. 
For the cross-correlation $\langle \dev{y}_1(0) \dev{y}_2(t)\rangle$
an analytical expression for $t'$ and its magnitude can be given:
\begin{align}
 \label{y1y2_t_min_eq}
          t'&=\frac{\tau}{2\mu}\ln\left( \frac{1+\mu}{1-\mu} \right), \\
 \label{y1y2min_eq}
         \langle \dev{y}_1(0) \dev{y}_2(t') \rangle&=\frac{1}{4}\left(
\left(\frac{1-\mu}{1+\mu}\right)^{\frac{1+\mu}{2\mu}} \hspace{-3mm} -
\left(\frac{1-\mu}{1+\mu}\right)^{\frac{1-\mu}{2\mu}} \right).
\end{align}

For finite values of the Weissenberg number $\Wi$ the cross-correlation
of the longitudinal displacements, $\langle \dev{x}_1(0) \dev{x}_2(t)\rangle $, also includes
contributions describing relaxation processes  perpendicular to $\vecd{q}_{12}$
with the eigenvalues $\lambda_2$ and $\lambda_4$. These additional contributions
cause larger shear-induced corrections to the cross\--corre\-lation in
eq.~(\ref{x1x2_0_eq})  than for the autocorrelation function in 
eq.~(\ref{x1x1_0_eq}), which can be seen by comparing the deviations between the 
dashed-dotted and the dashed curves in fig.~\ref{fig:03}
and in fig.~\ref{fig:06}.

\begin{figure}
  \begin{center}
\includegraphics[width=0.95\columnwidth]{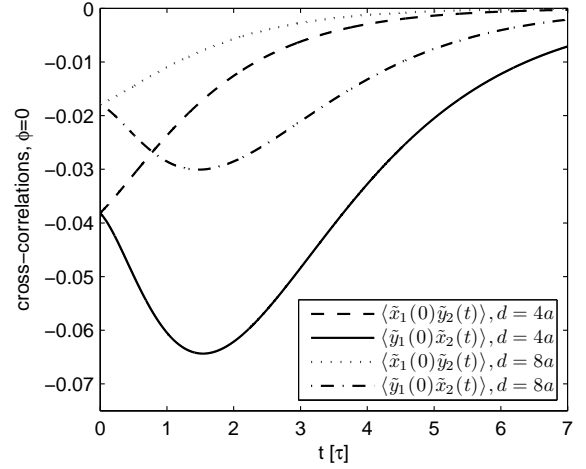}
  \end{center}
\caption{Shear-induced cross-correlations between orthogonal fluctuations of different particles
in the case, where $\vecd{q}_{12}$ is parallel to the streamlines, for $\Wi=1/2$ and different distances. 
}
\label{fig:07}
\end{figure}

For a single particle in a linear shear flow 
 shear-induced  cross-correlations 
between displacements 
along perpendicular directions in the shear plane were found,
which are proportional
to the Weissenberg number \cite{Holzer:2009.1}. 
In the two-particle system, one also obtains
cross-correlations between the displacements of two different
particles along orthogonal directions in the shear plane as described by: 
\begin{subequations}
 \label{xy2_0_eq}
\begin{align}
\label{x1y2_0_eq}
 &  \langle \dev{x}_1(0) \dev{y}_2(t) \rangle= \frac{\Wi}{4}\left( \frac{e^{-\lambda_2 t}}{2+3\mu}
- \frac{e^{-\lambda_4 t}}{2-3\mu} \right)\,,\\
\label{y2x1_0_eq}
 &  \langle \dev{y}_2(0) \dev{x}_1(t) \rangle= \frac{\Wi}{4\mu}\left( e^{-\lambda_2 t} +
e^{-\lambda_4 t} \right) \nonumber \\
 & \hspace{2mm} - \frac{\Wi}{2\mu}\left( \frac{(1+\mu)e^{-\lambda_1 t}}{2+3\mu} +
\frac{(1-\mu)e^{-\lambda_3 t}}{2-3\mu} \right)\,.
\end{align}
\end{subequations}
Both functions are plotted for different values of the trap distance in fig.~\ref{fig:07} and one can see that the magnitudes of the correlations are larger for smaller values of $d$.
Note that in general the particle index in the
previous equations can be interchanged: $\langle \dev{x}_1(0) \dev{y}_2(t) \rangle = \langle \dev{x}_2(0) \dev{y}_1(t) \rangle$.

The ratio between the two shear-induced cross\--corre\-lations 
at $t=0$ is independent of the Weissenberg number $\Wi$ and given by:
\begin{align}
 \label{x1y1x1y2rat_eq}
 \frac{\langle \dev{x}_1(0) \dev{y}_2(0) \rangle}{\langle \dev{x}_1(0) \dev{y}_1(0) \rangle}=\frac{-3\mu}{2}=\frac{-9a}{8d}\,.
\end{align}
This relation is reasonable for a sufficiently large ratio
$d/a$. It might serve in an experiment as a consistency check, since the bead
radius $a$ and the particle distance $d$ are usually well known.
If the two beads come close to each other, additional effects related
to their shear-induced rotation may come into play.
This is in general a limitation of our Langevin model (\ref{dgl_eq}) for point-like
particles with an effective hydrodynamic radius.
In Ref.~\cite{Stark:2004.1} particle rotations caused by an external torque have been
taken into account in a Langevin model for two trapped particles in a quiescent fluid.
This work describes a coupling between translational and rotational particle motions,
 which was confirmed by experiments \cite{Stark:2006.1}.

The asymmetry in eqs.~(\ref{xy2_0_eq}) with
respect to time, $t \to -t$,  has a similar origin 
as explained above for the shear-induced  one-particle correlations (\ref{xy1_0_eq}).
The location of the minimum is again mainly determined
by the relaxation time $\tau$ of the beads in the harmonic potentials.

Note, that for all described correlation functions the single particle case is recovered in the limit $d \to \infty$, which corresponds 
to $\mu \to 0$, for example:
\begin{align}
\label{y1x1lim_eq}
 \lim_{\mu \to 0} \langle \dev{y}_1(0) \dev{x}_1(t) \rangle = \frac{\Wi}{4}\left(1+2\frac{t}{\tau}\right)e^{-t/\tau}\,.
\end{align}

All the correlation functions presented in this subsection 
have already been measured in experiments \cite{Ziehl:2009.1,Ziehl:2010.1}.

\subsection{Perpendicular case: $\phi=\pi/2$}

For $\phi=\pi/2$ the time-dependence of the correlation functions and their magnitudes are changed compared to
the parallel orientation $\phi=0$. This subsection focuses on these differences. 

The formulas (\ref{ew3_eq}) for the eigenvalues of the matrix $\mat{M}$ were derived for the identity $\vecd{R}_0=\vecd{Q}$. This allows the determination of exact analytical formulas for the correlations.
In the present case, $\vecd{R}_{\pi/2}$ has to be determined numerically as well as the relaxation rates $\lambda_i$ and the exact functions ${\mat{C}}_{kl}(t)$. However, if we use the approximation $\vecd{R}_{\pi/2}=\vecd{Q}=d/2(0,1,0,0,-1,0)$, which is valid for small Weissenberg numbers, we obtain analytical formulas for the correlations that also describe quantitatively the characteristics of the functions ${\mat{C}_{kl}(t)}$ at large values of $\Wi$.
Within this assumption eqs.~(\ref{ew3_eq}) remain, but the eigenvalues exchange their meanings:
Now $\lambda_1$ and $\lambda_3$ belong to the longitudinal fluctuations in the $y$ direction, whereas $\lambda_2$ and $\lambda_4$ describe the transversal motions in $x$ and $z$ direction, 
perpendicular to $\vecd{q}_{12}$.
We also compare the numerical solutions with direct simulations of the Langevin equation (\ref{dgl_eq}).

\subsubsection{One-particle correlation}
\label{singleB}

\begin{figure}
  \begin{center}
\includegraphics[width=0.95\columnwidth]{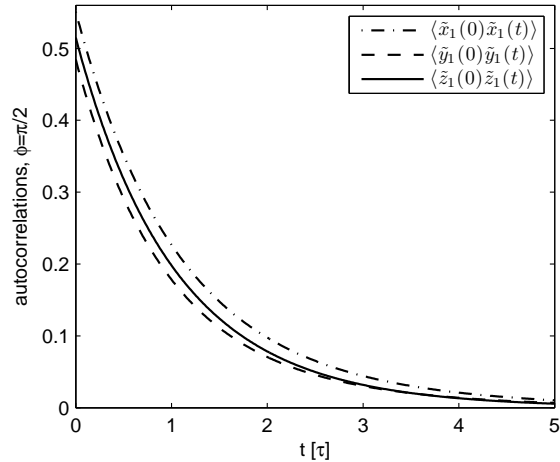}
  \end{center}
\caption{Autocorrelations for $\phi=\pi/2$, $d=4a$ and $\Wi=1/2$. $\langle
\dev{x}_1(0) \dev{x}_1(t)\rangle$, $\langle \dev{y}_1(0) \dev{y}_1(t)\rangle$ and $\langle \dev{z}_1(0)
\dev{z}_1(t)\rangle$ are different, because of the different relaxation rates and
additional $\Wi^2$-contributions in $x$ direction. One has  $\langle \dev{x}_1(0)
\dev{x}_1(t)\rangle=\langle \dev{z}_1(0) \dev{z}_1(t)\rangle$ only for $\Wi=0$.}
\label{fig:08}
\end{figure}

For the perpendicular configuration the one-particle autocorrelations 
are given 
for the approximation $\vecd{R}_{\pi/2}=\vecd{Q}$ by the following expressions:
\begin{subequations}
\label{xx_90_eq}
\begin{align}
\label{x1x1_90_eq}
  \langle \dev{x}_1(0) & \dev{x}_1(t) \rangle = \frac{1}{4}\left( e^{-\lambda_2 t} + e^{-\lambda_4
t} \right) \nonumber\\
 &  + \frac{\Wi^2}{4\mu}\left( \frac{(2\mu-1)~e^{-\lambda_4t}}{3\mu^2-5\mu+2} +
\frac{e^{-\lambda_3 t}}{2-3\mu} \right) \nonumber\\
 &  + \frac{\Wi^2}{4\mu}\left(\frac{(2\mu+1)e^{-\lambda_2 t}}{3\mu^2+5\mu+2}
-\frac{e^{-\lambda_1 t}}{2+3\mu}  \right)\,,\\
\label{y1y1_90_eq}
 \langle \dev{y}_1(0) & \dev{y}_1(t) \rangle =\frac{1}{4}\left(e^{-\lambda_1 t} + e^{-\lambda_3
t}\right)\,,   \\
\label{z1z1_90_eq}
  \langle \dev{z}_1(0) & \dev{z}_1(t) \rangle =\frac{1}{4}\left(e^{-\lambda_2 t} + e^{-\lambda_4
t}\right)\,.
\end{align}
\end{subequations}
While the correlations $\langle \dev{y}_1(0) \dev{y}_1(t) \rangle$
and $\langle \dev{z}_1(0) \dev{z}_1(t) \rangle$ of the displacements perpendicular to the
flow lines were identical in the case $\phi=0$, they are different
in the present case, because they describe positional fluctuations either parallel or perpendicular
to $\vecd{q}_{12}$. Only for a vanishing Weissenberg number one obtains $\langle \dev{x}_1(0) \dev{x}_1(t) \rangle=
\langle \dev{z}_1(0) \dev{z}_1(t) \rangle$, but for a finite shear rate 
the correlation functions are all different -- for the approximations in eqs.~(\ref{xx_90_eq}) 
as well as for the numerical solutions shown in fig.~\ref{fig:08}.

The shear-induced cross-correlations between particle displacements
along orthogonal directions  in the shear plane behave qualitatively
similar as in the case $\phi=0$. For the approximation $\vecd{R}_{\pi/2}=\vecd{Q}$ they are described by the
formulas
\begin{subequations}
\label{xy_90_eq}
\begin{align}
\label{x1y1_90_eq}
 \langle \dev{x}_1(0) &  \dev{y}_1(t) \rangle= \frac{\Wi}{4}\left( \frac{e^{-\lambda_1 t}}{2+3\mu}
+ \frac{e^{-\lambda_3 t}}{2-3\mu} \right)\,,\\
\label{y1x1_90_eq}
 \langle \dev{y}_1(0) &  \dev{x}_1(t) \rangle= \frac{\Wi}{4\mu}\left( e^{-\lambda_3 t} -
e^{-\lambda_1 t} \right) \nonumber \\
 & + \frac{\Wi}{2\mu}\left( \frac{(1+2\mu)e^{-\lambda_2 t}}{2+3\mu} -
\frac{(1-2\mu)e^{-\lambda_4 t}}{2-3\mu} \right)\,.
\end{align}
\end{subequations}
The magnitudes of the different contributions in eqs.~(\ref{xy_90_eq})
changed compared to the expressions in eqs.~(\ref{xy1_0_eq}),
which influences the positional probability distribution $P(\vecd{r})$ of one particle. 
The dependence of the inclination angle $\theta$ on the distance $d$ is weaker in the present case
than for $\phi=0$, see fig.~\ref{fig:13} in the next section.
Possible consequences of this
difference for the dynamics of beads-spring models for polymers in shear flows
are discussed in the concluding remarks.

\begin{figure}
  \begin{center}
  \includegraphics[width=0.95\columnwidth]{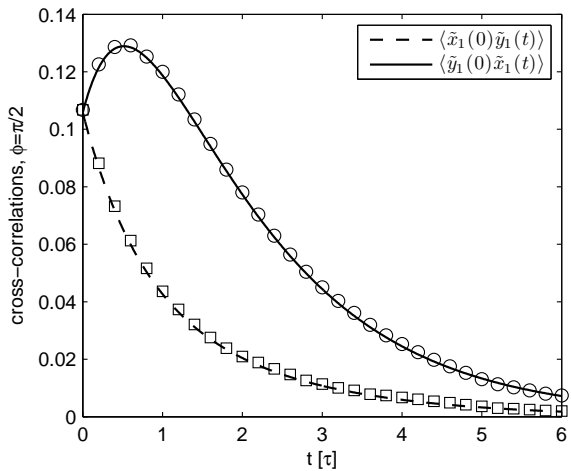}
  \end{center}
\caption{
Comparison of the numerically determined one-particle
cross-correlations (lines) with the results from the direct simulation of eq. (\ref{dgl_eq}) (circles, squares). The error bars are smaller than the symbols. Parameters: $d=4a, \Wi=1/2$.
}
\label{fig:09}
\end{figure}

The magnitudes of the correlations functions given by eqs.~(\ref{xy_90_eq})
are smaller than those determined numerically for the exact distance $\vecd{R}_{\pi/2}$,
which are plotted in fig.~\ref{fig:09}. In the same figure these exact solutions 
are compared with the results of a Brownian dynamics simulation of eq.~(\ref{dgl_eq}), where $2*10^6$ ensemble averages were made. This example illustrates the good agreement between both 
approaches, which is not affected by the choice of the Rotne-Prager tensor instead of the Oseen tensor in the
simulation.
The analytical expressions (\ref{xy_90_eq}) turn out to be a good approximation for the parameters $d \geq 8a$ and $\Wi \leq 0.1$.
Especially for smaller values of the trap distance $d$ deviations in the relaxation times occur.

\subsubsection{Inter-particle correlations}
\label{InterpartB}

\begin{figure}
  \begin{center}
\includegraphics[width=0.95\columnwidth]{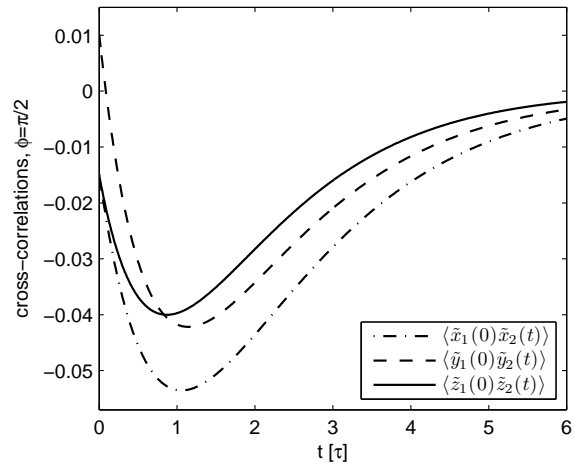}
  \end{center}
\caption{Inter-particle cross-correlations in the orthogonal case for $d=4a$ and $\Wi=1/2$. $\langle
\dev{x}_1(0) \dev{x}_2(t)\rangle$, $\langle \dev{y}_1(0) \dev{y}_2(t)\rangle$ and $\langle \dev{z}_1(0)
\dev{z}_2(t)\rangle$ are all different and anti-correlated with pronounced minima. 
One obtains $\langle \dev{x}_1(0) \dev{x}_2(t)\rangle=\langle \dev{z}_1(0)
\dev{z}_2(t)\rangle$ for $\Wi=0$.}
\label{fig:10}
\end{figure}

The cross-correlations of the random displacements between the
two particles in the same direction are given  for $\vecd{R}_{\pi/2}=\vecd{Q}$ by:
\begin{subequations}
\label{xy12_90_eq}
\begin{align}
\label{x1x2_90_eq}
   \langle \dev{x}_1(0) & \dev{x}_2(t) \rangle= \frac{1}{4}\left( e^{-\lambda_2 t} - e^{-\lambda_4
t} \right) \nonumber\\
 & + \frac{\Wi^2}{4\mu}\left( \frac{(1-2\mu)e^{-\lambda_4 t}}{3\mu^2-5\mu+2} -
\frac{e^{-\lambda_3 t}}{2-3\mu} \right) \nonumber\\
 & + \frac{\Wi^2}{4\mu}\left(\frac{(1+2\mu)e^{-\lambda_2 t}}{3\mu^2+5\mu+2}
-\frac{e^{-\lambda_1 t}}{2+3\mu}  \right)\,,\\
\label{y1y2_90_eq}
   \langle \dev{y}_1(0) & \dev{y}_2(t) \rangle=\frac{1}{4}\left(e^{-\lambda_1 t} - e^{-\lambda_3t}\right)\,,   \\
\label{z1z2_90_eq}
   \langle \dev{z}_1(0) & \dev{z}_2(t) \rangle=\frac{1}{4}\left(e^{-\lambda_2 t} - e^{-\lambda_4t}\right)\,.
\end{align}
\end{subequations}
Similar to the one-particle correlations in sect. \ref{singleB} and in contrast to the case $\phi=0$
in sect. \ref{interpartA}, all three
correlation functions are different for finite values of the Weissenberg number. This
is indicated by the expression in eqs.~(\ref{xy12_90_eq}) as well as by
the numerical correlation functions displayed in 
 fig.~\ref{fig:10}, and they are again all anti-correlated.

The most striking difference between the case $\phi=0$ and $\phi=\pi/2$ 
is found by looking at the shear-induced inter-particle correlations 
between orthogonal directions in the shear plane, which are given 
for $\vecd{R}_{\pi/2}=\vecd{Q}$ by the following expressions:
\begin{subequations}
 \label{xy2_90_eq}
\begin{align}
\label{x1y2_90_eq}
 \langle \dev{x}_1(0) &  \dev{y}_2(t) \rangle= \frac{\Wi}{4}\left( \frac{e^{-\lambda_1 t}}{2+3\mu}
- \frac{e^{-\lambda_3 t}}{2-3\mu} \right)\,,\\
\label{y1x2_90_eq}
 \langle \dev{y}_1(0) &  \dev{x}_2(t) \rangle= \frac{-\Wi}{4\mu}\left( e^{-\lambda_1 t} +
e^{-\lambda_3 t} \right) \nonumber \\
 &  + \frac{\Wi}{2\mu}\left( \frac{(1+2\mu)e^{-\lambda_2 t}}{2+3\mu} +
\frac{(1-2\mu)e^{-\lambda_4 t}}{2-3\mu} \right)\,.
\end{align}
\end{subequations}
A Taylor expansion of the correlation 
function $ \langle \dev{y}_1(0)  \dev{x}_2(t) \rangle$ 
 with respect to
 small values of $\mu$ reveals the difference between the
parallel and the perpendicular orientations.
 For $\phi=0$ one obtains up to the linear order of $\mu$,
\begin{align}
\label{ty1x2_00_eq}
 \langle \dev{y}_1(0)  \dev{x}_2(t) \rangle \sim \mu\left(3+4\frac{t}{\tau}+6\frac{t^2}{\tau^2} \right)e^{-t/\tau}\,,
\end{align}
whereas for $\phi=\pi/2$ one obtains,
\begin{align}
\label{ty1x2_90_eq}
 \langle \dev{y}_1(0)  \dev{x}_2(t) \rangle \sim \mu\left(3+2\frac{t}{\tau}+6\frac{t^2}{\tau^2} \right)e^{-t/\tau}\,.
\end{align}
The first expression has only {\it one} extremum as a function of time.
The different prefactor of $te^{-t/\tau}$ in the second expression 
is the origin of an additional extremum in the case $\phi=\pi/2$.

\begin{figure}
  \begin{center}
\includegraphics[width=0.95\columnwidth]{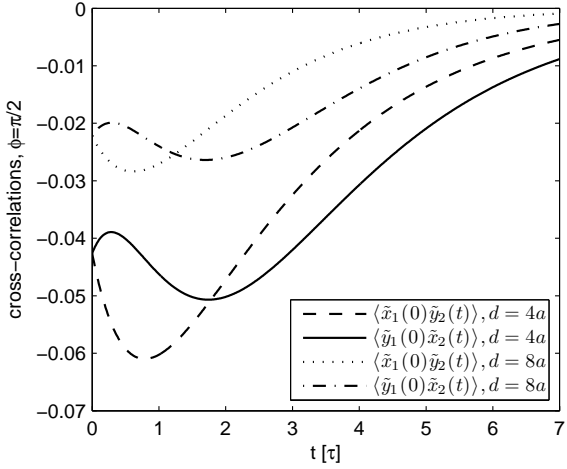}
  \end{center}
\caption{
Shear-induced inter-particle correlations in case of a tilted trap
for a Weissenberg number $\Wi=1/2$ and for the two distances $d=4a$ and $d=8a$. The time-dependence is different from the case $\phi=0$, cf. Fig.~\ref{fig:07}.
}
\label{fig:11}
\end{figure}

For $\vecd{Q}=d/2(0,1,0,0,-1,0)$ the angle between the vector connecting
the resulting mean positions of the particles, $\langle \vecd{r}_{12} \rangle$, and the $y$ axis increases as a function of $\Wi$.
In this configuration the numerical solution for $\langle \dev{y}_1(0)  \dev{x}_2(t) \rangle$
does not show a second extremum. 
However, if $\vecd{Q}$ is tilted against the flow direction, like
\begin{align}
 \label{tilted_eq}
\vecd{Q}=\frac{d}{2\sqrt{1+\Wi^2}}(-\Wi,1,0,\Wi,-1,0)\,,
\end{align}
the resulting vector $\langle \vecd{r}_{12} \rangle$ becomes nearly parallel to the $y$ axis and the second
extremum of $\langle \dev{y}_1(0)  \dev{x}_2(t) \rangle$ is obtained again, as
predicted by the approximation in eq.~(\ref{ty1x2_90_eq}).
The result for the tilted trap is shown in fig.~\ref{fig:11} for two different values of $d$.
Moreover, the extrema
are now stronger pronounced than predicted by eq.~(\ref{ty1x2_90_eq}).
Comparing this with fig.~\ref{fig:07}, where the results for $\phi=0$  are plotted, it can be clearly seen,
that the correlation $\langle \dev{y}_1(0) \dev{x}_2(t) \rangle$ now has an additional local maximum,
cf. solid and dash-dotted line in fig.~\ref{fig:07}, 
and that $\langle \dev{x}_1(0) \dev{y}_2(t) \rangle$ has a minimum at small values of the time $t$,
cf. dashed and dotted lines in fig.~\ref{fig:07}.

\subsection{Oblique case: $\phi=\pi/4$}\label{sec: oblique}

If the connection vector $\vecd{q}_{12}$ between the two potential minima
is oblique to the flow direction, further aspects for the correlation functions may come
into play.
In principle, the matrix ${\mat{C}_{kl}}(t)$ can be calculated for any angle $\phi$. 
However, we focus on the special case $\phi=\pi/4$ as an example.
This has the advantage that analytical expressions can be obtained
under the assumption $\vecd{R}_{\pi/4}=\vecd{Q}$.
These approximate solutions remain rather compact and may serve as a guide
for the qualitative behavior of the correlation functions.

Since the linear shear flow (\ref{u_eq}) dictates a preferred direction in the system,
the particle fluctuations are decomposed into
eigenmodes parallel and perpendicular to the $x$ axis as described in sect.~\ref{sec: calc}.
That's why in the oblique configuration the longitudinal displacements consist of a superposition of the eigenmodes in the shear plane. 
If $\phi=\pi/4$, the eigenvalues of the matrix ${\mat{M}}$ are given for 
 $\vecd{R}_{\pi/4}=\vecd{Q}=d\sqrt{2}/4(1,1,0,-1,-1,0)$ by the following expressions,
\begin{subequations}
\begin{align}
  \lambda_1&=\frac{2+3\mu+\sqrt{\mu^2-2\mu \Wi}}{2\tau}\,, \\
  \lambda_2&=\frac{2+3\mu-\sqrt{\mu^2-2\mu \Wi}}{2\tau}\,, \\
  \lambda_3&=\frac{1+\mu}{\tau}\,,\\ 
  \lambda_4&=\frac{2-3\mu-\sqrt{\mu^2+2\mu \Wi}}{2\tau}\,, \\
  \lambda_5&=\frac{2-3\mu+\sqrt{\mu^2+2\mu \Wi}}{2\tau}\,, \\
  \lambda_6&=\frac{1-\mu}{\tau}\,.
 \end{align}
\label{ewo_eq}
\end{subequations}
While $\lambda_1$ and $\lambda_4$ describe the parallel and anti-parallel relaxation modes in the $x$ direction,
$\lambda_2$ and $\lambda_5$ belong to the $y$ direction, and 
$\lambda_3$ and $\lambda_6$ to the $z$ direction.
In contrast to the parallel case, the relaxation processes 
along the $x$ and $y$ direction in the shear plane are now affected by the shear rate.
So the corresponding eigenvalues depend directly on the Weissenberg number.
This was not the case in the previous section for $\vecd{R}_{\pi/2}=\vecd{Q}$.
As long as $\Wi \neq 0$, the six relaxation parameters are different and can partly become complex numbers,
causing oscillatory contributions to the functions ${{\mat{C}_{kl}}(t)}$.

In order to write down the full expressions for the correlations in a compact form, we introduce the following notions similar to eq.~(\ref{ckl_eq}):
\begin{subequations}
\begin{align}
 &g_{1,1}=\mu +\frac{\Wi(\lambda_1-\lambda_2)}{2\lambda_1},\\
	&g_{2,1}=\mu - \frac{\Wi(\lambda_1-\lambda_2)}{2\lambda_2},\\
 &g_{4,1}=\mu +\frac{\Wi(\lambda_5-\lambda_4)}{2\lambda_4}, \\
	&g_{5,1}=\mu -\frac{\Wi(\lambda_5-\lambda_4)}{2\lambda_5}, \\
 &g_{1,2}=\mu -\frac{\Wi(3\lambda_1+\lambda_2)}{2\lambda_1}, \\
	&g_{2,2}=\mu - \frac{\Wi(3\lambda_2+\lambda_1)}{2\lambda_2},\\
 &g_{4,2}=\mu +\frac{\Wi(3\lambda_4+\lambda_5)}{2\lambda_4}, \\
	&g_{5,2}=\mu +\frac{\Wi(3\lambda_5+\lambda_4)}{2\lambda_5}.
\end{align}
\end{subequations}

Analogous to the previous subsections we discuss the one-particle correlations first.
The autocorrelations of a single particle show the same behavior as in the case
$\phi=\pi/2$. They are different in distinct directions as long as $\Wi \neq 0$ and decay exponentially in time.
For $\vecd{R}_{\pi/4}=\vecd{Q}$ the analytical formulas read
\begin{subequations}
 \label{xx1_45_eq}
\begin{align}
\label{x1x1_45_eq}
 \langle \dev{x}_1(0) & \dev{x}_1(t) \rangle= \frac{g_{1,1}e^{-\lambda_1 t}}{8\mu} + \frac{g_{2,1}e^{-\lambda_2 t}}{8\mu} \nonumber\\&+ \frac{g_{4,1}e^{-\lambda_4 t}}{8\mu} + \frac{g_{5,1}e^{-\lambda_5 t}}{8\mu} \,,\\
\label{y1y1_45_eq}
 \langle \dev{y}_1(0) & \dev{y}_1(t) \rangle= \frac{g_{1,2}e^{-\lambda_1 t}}{8(\mu-2\Wi)} + \frac{g_{2,2}e^{-\lambda_2 t}}{8(\mu-2\Wi)} \nonumber\\&+ \frac{g_{4,2}e^{-\lambda_4 t}}{8(\mu+2\Wi)} + \frac{g_{5,2}e^{-\lambda_5 t}}{8(\mu+2\Wi)} \,, \\
\label{z1z1_45_eq}
\langle \dev{z}_1(0) & \dev{z}_1(t) \rangle= \frac{1}{4}\left(e^{-\lambda_3 t} + e^{-\lambda_6 t}\right)\,.
\end{align}
\end{subequations}
However, in contrast to eqs.~(\ref{xx_90_eq})
one can see that the two autocorrelation functions for the particle displacements in the shear plane, namely $\langle \dev{x}_1(0)  \dev{x}_1(t) \rangle$ and $\langle \dev{y}_1(0)  \dev{y}_1(t) \rangle$, include the four corresponding relaxation modes and depend in a complex way on the Weissenberg number.

\begin{figure}
  \begin{center}
\includegraphics[width=0.98\columnwidth]{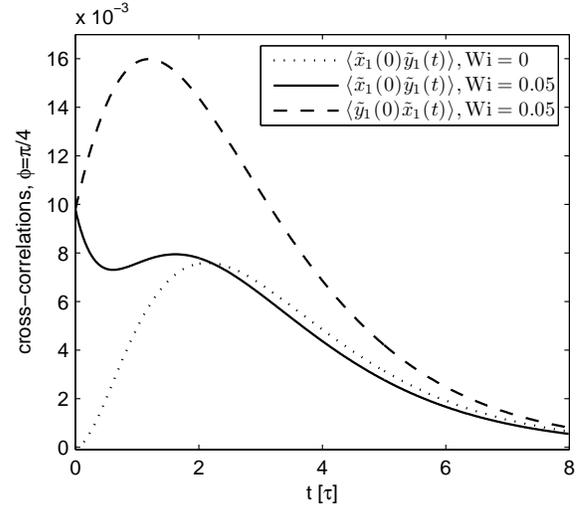}
  \end{center}
\caption{Single-particle correlations between orthogonal displacements in the shear plane for different Weissenberg numbers, $\phi=\pi/4$ and $d=4a$.
In contrast to the parallel or the perpendicular case, $\langle \dev{x}_1(0) \dev{y}_1(t) \rangle$ and $\langle \dev{y}_1(0) \dev{x}_1(t) \rangle$ remain finite for $\Wi=0$.}
\label{fig:12}
\end{figure}

The one-particle cross-correlations between orthogonal positional fluctuations
in the shear plane, like $\langle \dev{x}_1(0)  \dev{y}_1(t) \rangle$, are purely shear-induced in the parallel and in the
perpendicular case.
They are linear functions of the parameter $\Wi$, which means, that they vanish in the limit of zero shear rate.
This is different in the orthogonal configuration.
For $\phi=\pi/4$ and $\vecd{R}_{\pi/4}=\vecd{Q}$ the correlation functions are given by the expressions:

\begin{subequations}
\label{xy1_45_eq}
\begin{align}
\label{x1y1_45_eq}
 \langle \dev{x}_1(0) & \dev{y}_1(t) \rangle= \frac{g_{1,1}e^{-\lambda_1 t}}{8\sqrt{\mu^2-2\mu \Wi}} - \frac{g_{2,1}e^{-\lambda_2 t}}{8\sqrt{\mu^2-2\mu \Wi}} \nonumber\\&+ \frac{g_{4,1}e^{-\lambda_4 t}}{8\sqrt{\mu^2+2\mu \Wi} } - \frac{g_{5,1}e^{-\lambda_5 t}}{8\sqrt{\mu^2+2\mu \Wi}} \,,\\
\label{y1x1_45_eq}
 \langle \dev{y}_1(0) & \dev{x}_1(t) \rangle= \frac{g_{1,2}e^{-\lambda_1 t}}{8\sqrt{\mu^2-2\mu \Wi }} - \frac{g_{2,2}e^{-\lambda_2 t}}{8\sqrt{\mu^2-2\mu \Wi }} \nonumber\\&+ \frac{g_{4,2}e^{-\lambda_4 t}}{8\sqrt{\mu^2+2\mu \Wi }} - \frac{g_{5,2}e^{-\lambda_5 t}}{8\sqrt{\mu^2+2\mu \Wi }} \,\,.
\end{align}
\end{subequations}
In the limit of very large particle distances, when $\mu \to 0$, $\langle \dev{x}_1(0) \dev{y}_1(t) \rangle$
and $\langle \dev{y}_1(0)  \dev{x}_1(t) \rangle$ resemble the one-particle case given by eq.~(\ref{y1x1lim_eq}), whereas in the limit $\Wi \to 0$ the two functions become equal but do not vanish in contrast to the previous subsections.
Instead of the approximate expressions (\ref{xy1_45_eq})
we display in fig.~\ref{fig:12} the full numerical solution for these correlation functions.
If $\Wi=0$, the identity $ \langle \dev{x}_1(0)  \dev{y}_1(t) \rangle=\langle \dev{y}_1(0)  \dev{x}_1(t) \rangle$ is obtained, cf. dotted line in fig.~\ref{fig:12}.
For larger values of $\Wi$ the two functions resemble the curves shown in fig.~\ref{fig:04}.

\begin{figure}
  \begin{center}
\includegraphics[width=0.98\columnwidth]{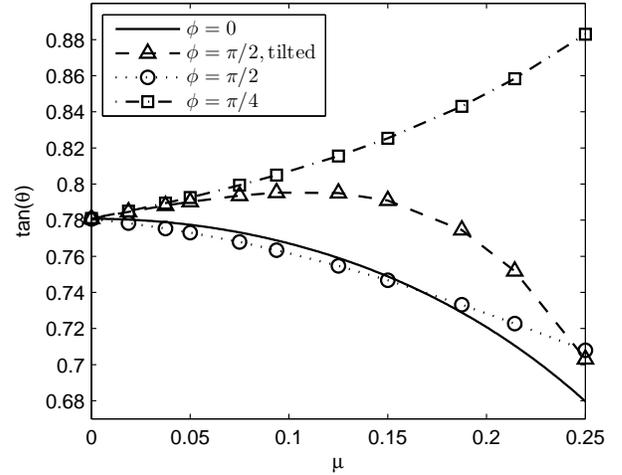}
  \end{center}
\caption{The inclination angle of the single particle distribution, as given by eq.~(\ref{tanalpha_eq}), is shown as a function of the parameter $\mu$. For $\Wi=1/2$ the exact results for the different setups described in the text are compared.}
\label{fig:13}
\end{figure}

According to eq.~(\ref{tanalpha_eq}) and eq.~(\ref{tanG}) the inclination angle $\theta$ of the elliptical positional probability distribution is determined by the static single-particle correlation functions, which depend on the trap distance $d$.
In fig.\ref{fig:13} $\tan(\theta)$ is shown as a function of the parameter $\mu=3a/(4d)$ for different trap configurations.
The analytical expression from the parallel case (solid line) is compared with the full numerical solutions for the other setups. For $\phi=\pi/2$ (circles) the monotonous decrease of $\tan(\theta)$ is weaker than for $\phi=0$. For the tilted configuration given by eq.~(\ref{tilted_eq}) $\theta$ has a broad maximum (triangles), whereas in the oblique case, $\phi=\pi/4$, the inclination angle is increasing continuously with increasing $\mu$ (squares).

\begin{figure}
  \begin{center}
\includegraphics[width=0.98\columnwidth]{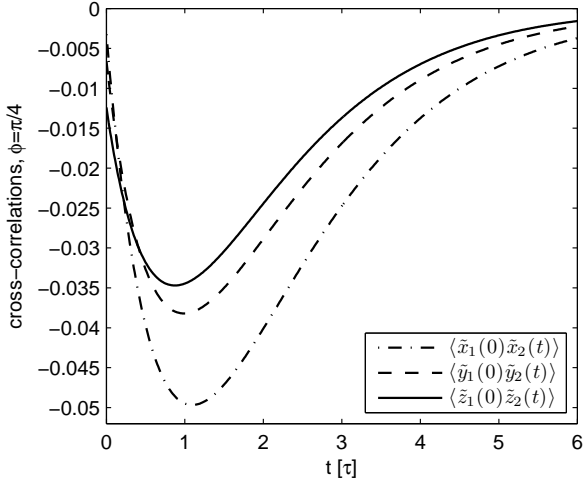}
  \end{center}
\caption{Inter-particle cross-correlations in the oblique case for $d=4a$ and $ \Wi=1/2$. In the limit $\Wi=0$, $\langle \dev{x}_1(0) \dev{x}_2(t)\rangle$ 
and $\langle \dev{y}_1(0) \dev{y}_2(t)\rangle$ become equal.}
\label{fig:14}
\end{figure}

The inter-particle cross-correlations in parallel directions are again anti-correlated in the present case. 
For the approximation $\vecd{R}_{\pi/4}=\vecd{Q}$, we obtain:
\begin{subequations}
 \label{xx2_45_eq}
\begin{align}
\label{x1x2_45_eq}
 \langle \dev{x}_1(0) & \dev{x}_2(t) \rangle= \frac{g_{1,1}e^{-\lambda_1 t}}{8\mu} + \frac{g_{2,1}e^{-\lambda_2 t}}{8\mu} \nonumber\\&- \frac{g_{4,1}e^{-\lambda_4 t}}{8\mu} - \frac{g_{5,1}e^{-\lambda_5 t}}{8\mu} \,,\\
\label{y1y2_45_eq}
 \langle \dev{y}_1(0) & \dev{y}_2(t) \rangle= \frac{g_{1,2}e^{-\lambda_1 t}}{8(\mu-2\Wi)} + \frac{g_{2,2}e^{-\lambda_2 t}}{8(\mu-2\Wi)} \nonumber\\&- \frac{g_{4,2}e^{-\lambda_4 t}}{8(\mu+2\Wi)} - \frac{g_{5,2}e^{-\lambda_5 t}}{8(\mu+2\Wi)} \,, \\
\label{z1z2_45_eq}
\langle \dev{z}_1(0) & \dev{z}_2(t) \rangle= \frac{1}{4}\left(e^{-\lambda_3 t} - e^{-\lambda_6 t}\right)\,.
\end{align}
\end{subequations}
The eqs.~(\ref{xx2_45_eq}) indicate that these correlations are different from each other, similar to the perpendicular case. The correlation functions for the correct $\vecd{R}_{\pi/4}$ are shown in fig.~\ref{fig:14}.

\begin{figure}
  \begin{center}
\includegraphics[width=0.98\columnwidth]{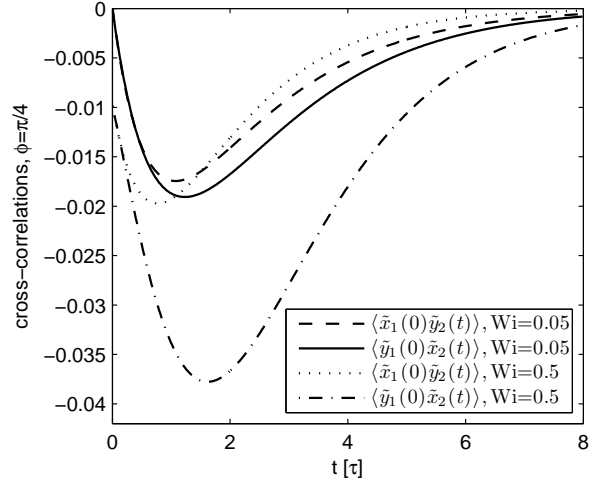}
  \end{center}
\caption{Cross-correlations in the oblique case for $d=4a$ and different Weissenberg numbers.
The amplitude of $\langle \dev{y}_1(0) \dev{x}_2(t) \rangle$ is increasing much stronger with increasing values of $\Wi$ than the amplitude of $\langle \dev{x}_1(0) \dev{y}_2(t) \rangle$.}
\label{fig:15}
\end{figure}

The cross-correlations $\langle \dev{x}_1(0)  \dev{y}_2(t) \rangle$ and $\langle \dev{y}_1(0)  \dev{x}_2(t) \rangle$ show the strongest dependence on the orientation of the trapped particles with respect to the flow direction. 
If $\phi=\pi/4$, they have one minimum each, which is different from the case $\phi=0$ and from the tilted configuration (\ref{tilted_eq}). Within the approximation, the functions are given by:
\begin{subequations}
 \label{xy2_45_eq}
\begin{align}
\label{x1y2_45_eq}
 \langle \dev{x}_1(0) & \dev{y}_2(t) \rangle= \frac{g_{1,1}e^{-\lambda_1 t}}{8\sqrt{\mu^2-2\mu \Wi}} - \frac{g_{2,1}e^{-\lambda_2 t}}{8\sqrt{\mu^2-2\mu \Wi}} \nonumber\\&- \frac{g_{4,1}e^{-\lambda_4 t}}{8\sqrt{\mu^2+2\mu \Wi}} + \frac{g_{5,1}e^{-\lambda_5 t}}{8\sqrt{\mu^2+2\mu \Wi}} \,,\\
\label{y2x1_45_eq}
 \langle \dev{y}_1(0) & \dev{x}_2(t) \rangle= \frac{g_{1,2}e^{-\lambda_1 t}}{8\sqrt{\mu^2-2\mu \Wi}} - \frac{g_{2,2}e^{-\lambda_2 t}}{8\sqrt{\mu^2-2\mu \Wi}} \nonumber\\&- \frac{g_{4,2}e^{-\lambda_4 t}}{8\sqrt{\mu^2+2\mu \Wi}} + \frac{g_{5,2}e^{-\lambda_5 t}}{8\sqrt{\mu^2+2\mu \Wi}} \,.
\end{align}
\end{subequations}
As illustrated by the numerical results for the correct $\vecd{R}_{\pi/4}$ in fig.~\ref{fig:15}, it is remarkable that $\langle \dev{x}_1(0)  \dev{y}_2(t) \rangle$ does not change very much with the Weissenberg number, while the amplitude of $\langle \dev{y}_1(0)  \dev{x}_2(t) \rangle$ increases with $\Wi$.
Their asymptotic behavior for $\Wi \to 0$ is similar to the single-particle correlations given by eqs.~(\ref{xy1_45_eq}).

\section{Conclusion}\label{sec: conclusion}

In this work we investigated the dynamics of two Brownian particles,
each trapped by a harmonic potential and exposed to a linear shear flow.
The one-particle and the inter-particle positional correlation functions, which can be measured in an experiment, were calculated by solving an appropriate Langevin model.
We discussed the correlations in detail as a function of the distance between the two
traps, as a function of the Weissenberg number $\Wi$, and for three different configurations, where
the vector connecting the two potential minima was either parallel, perpendicular,
or oblique with respect to the flow direction. This relative orientation strongly affects the time-dependence
of the correlation functions.
For the parallel configuration exact analytical expressions were presented. Otherwise, we provided numerical solutions and analytical approximations for the correlation functions.

Although the stochastic forces in our model were assumed to be uncorrelated along orthogonal directions, we found a coupling between perpendicular particle displacements caused by the shear flow, similar to that in Ref.~\cite{Holzer:2009.1}.
The resulting shear-induced cross-correlations depend linearly on the Weissenberg number, $\Wi$, and occur also between orthogonal fluctuations of different particles.
These inter-particle correlations have zero, one, or two extrema as a function of time, depending on the particle configuration.
Besides generating new cross-correlations, the shear flow causes a contribution proportional to $\Wi^2$ in the
correlation functions of particle fluctuations along parallel directions.

Due to the hydrodynamic interaction between the two particles
the magnitudes of the one-particle correlations are enhanced with decreasing trap distance $d$, while
in the limit of large distances the single-particle results presented in Ref.~\cite{Holzer:2009.1} were recovered.
Moreover, we found a significant impact of the parameter $d$ on the positional 
probability distribution of each particle in the shear flow. 
The shape of the elliptical distribution
is tilted and stretched when the two particles approach each other in the parallel
configuration. The same effect is observed
when the shear rate is increased. So the shear-flow effects are enhanced by the presence 
of a second particle.
In the oblique configuration the opposite effect is observed.

The shear-induced cross-correlations investigated in 
this work are of the same origin as the correlations between orthogonal fluctuations
of a single freely floating particle as discussed 
in Ref.~\cite{Bedeaux:1995.1,Drossinos:05}. 
However, if the Brownian particles are trapped,
a direct experimental detection of these correlations
becomes possible.
This has been achieved recently in Ref.~\cite{Ziehl:2009.1}, where
two polystyrene latex spheres were trapped 
by optical tweezers and exposed to a linear shear flow in a special microfluidic device.
With this setup the cross-correlations of the positional fluctuations
for the parallel case, as shown in fig.~\ref{fig:07},
were measured directly for the first time. Moreover,
we predict an additional extremum for the shear-induced
inter-particle correlations, if the two particles are trapped 
perpendicular to the flow lines, cf. fig.~\ref{fig:11}.

The two hydrodynamically interacting beads are treated
as point-particles. In forthcoming works, the presented
results on the shear-induced correlations are extended
by taking the finite particle extension into account.
Preliminary investigations show that the effects of particle
rotations do not change the major trends in this work \cite{Schreiber:2010.2},
and for small $\Wi$ the rotation can be neglected anyhow.

The present article provides insight on the Brownian motion
of two trapped particles that might be useful for the analysis
of the stochastic dynamics of a bead-spring model for polymers too, where the 
Brownian particles are connected along a chain and fluctuate around some mean 
distance to their next neighbors.
The fluctuations along and perpendicular to the connection vector between two neighboring
beads may exhibit similar correlations as for the three model configurations investigated in this work, cf.
fig.~\ref{fig:01}.
In this spirit, a profound analysis of the stochastic motion of a bead-spring
model in a linear shear flow is an interesting task and may contribute further to the understanding of polymer dynamics.

{\it {Acknowledgments.-}}
We would like to thank C. Wagner, A. Ziehl, S. Schreiber and D. Kienle for instructive discussions
and the anonymous referee for very useful questions and suggestions.
This work has been supported by the German Science Foundation (DFG) through
the priority program on nano- and microfluidics SPP 1164 and by the Bayerisch-Franz\"osisches
Hochschulzentrum (BFHZ).

%
%

\end{document}